\documentclass{acm_proc_article-sp}

\usepackage[english]{babel}
\usepackage[T1]{fontenc}

\usepackage[a4paper,top=3cm,bottom=2cm,left=3cm,right=3cm,marginparwidth=1.75cm]{geometry}

\usepackage{amsmath}
\usepackage{graphicx}
\usepackage[colorinlistoftodos]{todonotes}

\usepackage[utf8]{inputenc}
\usepackage{amssymb}
\usepackage{mathtools}
\usepackage{enumerate}
\usepackage{bbm}
\usepackage{subcaption}
\usepackage{grffile}

\graphicspath{{figures/}}

\title{Ground Control to Major Tom: the importance of field surveys in remotely sensed data analysis}
\author{Ian Bolliger, Tamma Carleton, Solomon Hsiang, Jonathan Kadish, \\  Jonathan Proctor, Benjamin Recht, Esther Rolf, Vaishaal Shankar}

\numberofauthors{8}
\author{
\alignauthor
Ian W. Bolliger\\
       \affaddr{University of California, Berkeley}\\
       \affaddr{Berkeley, CA}\\
       \email{bolliger@berkeley.edu}
\alignauthor
Tamma Carleton\\
       \affaddr{University of California, Berkeley}\\
       \affaddr{Berkeley, CA}\\
       \email{tcarleton@berkeley.edu}
\alignauthor
Solomon Hsiang\\
       \affaddr{University of California, Berkeley}\\
       \affaddr{Berkeley, CA}\\
       \email{shsiang@berkeley.edu}
\and
\alignauthor
Jonathan Kadish\\
       \affaddr{University of California, Berkeley}\\
       \affaddr{Berkeley, CA}\\
       \email{jkadish@berkeley.edu}
\alignauthor
Jonathan Proctor\\
       \affaddr{University of California, Berkeley}\\
       \affaddr{Berkeley, CA}\\
       \email{proctor@berkeley.edu}
\alignauthor
Benjamin Recht\\
       \affaddr{University of California, Berkeley}\\
       \affaddr{Berkeley, CA}\\
       \email{brecht@berkeley.edu}
\and
\alignauthor
Esther Rolf\\
       \affaddr{University of California, Berkeley}\\
       \affaddr{Berkeley, CA}\\
       \email{esther\_rolf@berkeley.edu}
\alignauthor
Vaishaal Shankar\\
       \affaddr{University of California, Berkeley}\\
       \affaddr{Berkeley, CA}\\
       \email{vaishaal@berkeley.edu}
}

\begin{document}
\maketitle

\section{Background}

Recent increases in the availability of remotely sensed data have created a remarkable opportunity to measure fundamental indicators of human well-being at a global scale. These new data sources, such as satellite-derived climate variables and remote sensing imagery, are particularly valuable in locations where direct on-site data are rarely collected, such as developing countries and regions affected by conflict. Increasingly high-resolution imagery provides rich information on a host of landscape characteristics that may be correlated with various social and economic outcomes of policy and research interest. A growing set of approaches is being developed to condense data contained within these images into meaningful and comparable metrics for prediction of socioeconomic and development indicators. These advances open a window to the construction and analysis of globally-comprehensive datasets that shed light on factors of living conditions previously outside the reach of standard data collection efforts. Such approaches for transforming imagery into meaningful social data have been designed to predict outcomes such as: socioeconomic class in Lima, Peru \cite{tapiador_deriving_2011}; quality of life in cities in Indiana \cite{li_measuring_2007,jensen_using_2005} and Georgia \cite{lo_integration_1997}; and most recently, household consumption and assets in five African countries \cite{jean_combining_2016}. 

These existing remote sensing approaches are highly specialized; researchers develop application-specific techniques to predict socioeconomic outcomes in a location of interest, relying on available observational data. This generates two key limitations. First, ground truth data are often sparse in places where remote sensing tools are of unique value. The predictive skill of these techniques is constrained by available data, and returns on increases to the quantity or quality of observational data remain unknown. Second, findings are heavily dependent on the method used to transform imagery into usable image features for prediction. For example, remote sensing imagery applications often treat the pixel as the observational unit, but recent advances in machine learning algorithms and computational power have enabled information on the spatial structure of an entire image to be extracted \cite{li_review_2014}.

In this project, we build a modular, scalable system that can collect, store, and process millions of satellite images. We test the relative importance of both of the key limitations constraining the prevailing literature by applying this system to a data-rich environment. To overcome classic data availability concerns, and to quantify their implications in an economically meaningful context, we operate in a data rich environment and work with an outcome variable directly correlated with key indicators of socioeconomic well-being. We collect public records of sale prices of homes within the United States, and then gradually degrade our rich sample in a range of different ways which mimic the sampling strategies employed in actual survey-based datasets. Pairing each house with a corresponding set of satellite images, we use image-based features to predict housing prices within each of these degraded samples. To generalize beyond any given featurization methodology, our system contains an independent featurization module, which can be interchanged with any preferred image classification tool. 
We test this system under five realistic sampling strategies, and with three image classification techniques. Our initial findings demonstrate that while satellite imagery can be used to predict housing prices with considerable accuracy, the size and nature of the ground truth sample is a fundamental determinant of the usefulness of imagery for this category of socioeconomic prediction. We quantify the returns to improving the distribution and size of observed data, and show that the image classification method is a second-order concern. Our results provide clear guidance for the development of adaptive sampling strategies in data-sparse locations where satellite-based metrics may be integrated with standard survey data, while also suggesting that advances from image classification techniques for satellite imagery could be further augmented by more robust sampling strategies.

\section{Our approach}

\begin{figure*}[!h]
\centering
\includegraphics[width=\textwidth]{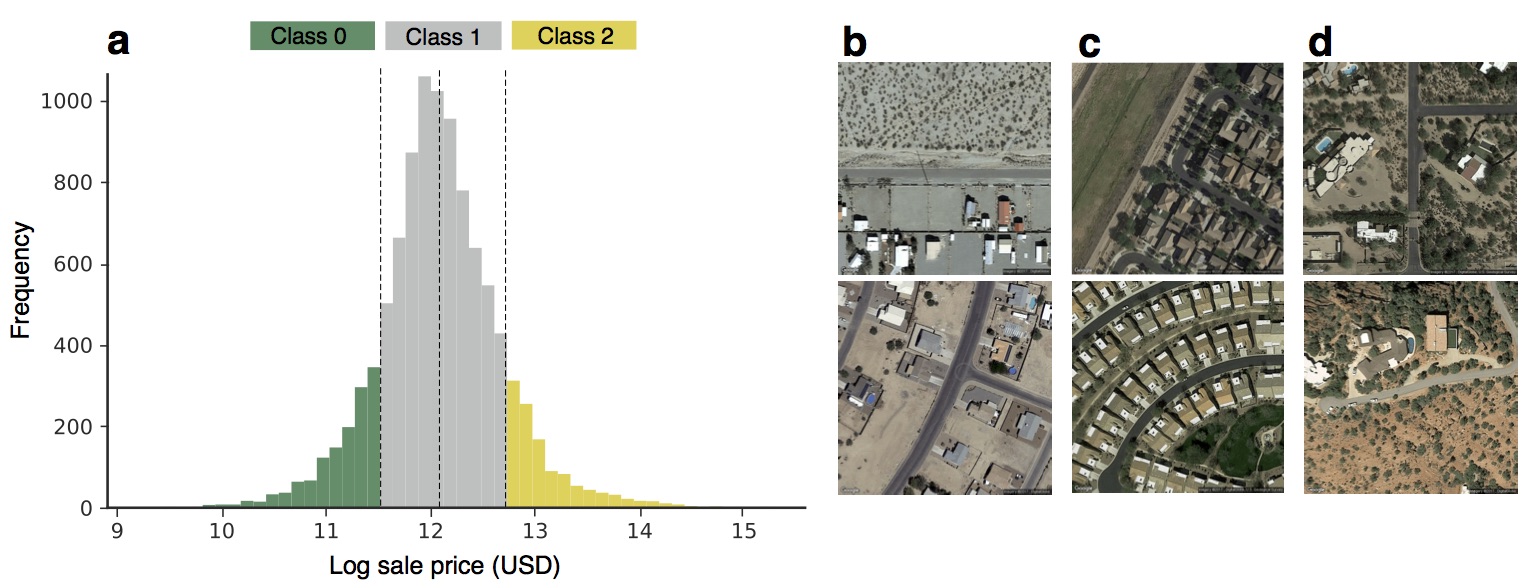}
\caption{\textbf{(a)} Distribution of home sale prices in Arizona. Examples of satellite images which are featurized to predict average sale price, representing houses from: \textbf{(b)} the poorest $\sigma$ of the housing price distribution (class 0); \textbf{(c)} the middle of the housing price distribution (class 1); and \textbf{(d)} the richest $\sigma$ of the housing price distribution (class 2).
\label{fig:distribution}}
\end{figure*}

Wealth estimation is a widely applicable and generally difficult problem in remote sensing \cite{jean_combining_2016, henderson2012measuring}. To execute this task in a data-rich setting, we use a dataset of all home sales in Arizona from 2010 to 2016, pulling georeferenced data from public records and pairing each home with associated aerial satellite imagery. We link home prices with images by associating every satellite image with the average log price of homes which fall within the image. To capture meaningful variation in housing prices while keeping the model simple, we classify a satellite image into one of three regions of the Arizona home price distribution (see Fig. \ref{fig:distribution}).

We simulate real world data collection processes that are common in data-scarce regions by sub-sampling our dataset of all home prices under various sampling strategies (see Fig. \ref{fig:uar_matrix}(m:p)). These sampling strategies are designed to reflect how actual surveys are conducted. For example, the Demographic and Health Survey (DHS) and the Living Standards Measurement Survey (LSMS) both survey individuals using a cluster-based survey approach (we call this ``cluster'' sampling). Administrative borders with differential data collection efforts on either side often generate samples with abrupt geographic boundaries (we call this ``latitudinally stratified'' or ``longitudinally stratified'' sampling). Using each of these sampling strategies, we then compare prediction performance of models trained on each sub-sampled dataset. The strategies we consider for this work are: 

\begin{enumerate}
\item Uniform at random (UAR) sampling.  Samples are collected uniformly at random from the distribution of homes in the area of interest (AOI). 
\item Cluster sampling. Samples are collected from a pre-specified number of clusters geographically spread out in the AOI.
\item Latitudinally stratified sampling. Samples are collected north or south of a particular latitude line, but uniform across longitudes in the AOI.
\item Longitudinally stratified sampling. Samples are collected west or east of a particular longitude line, but uniform across latitudes in the AOI.
\end{enumerate}

Image classification is a well studied problem in computer vision, and the current state-of-the-art algorithms for this task are Convolutional Neural Networks (CNNs). To generalize our findings beyond a single classification technique, we use three distinct baseline approaches:
\begin{enumerate}
\item GIST descriptor, a low dimensional representation of images specialized for ``scene'' descriptions \cite{oliva2001modeling}.

\item Pre-trained CNN (the VGG network), trained on a data set of 1.2 million natural images, from 1000 categories  The penultimate layer of this CNN provides a succinct representation of images that works well for many classification tasks \cite{simonyan2014very, huh2016makes, oquab2014learning}.
\item Random CNN (Coates-Ng). Random convolutional networks, for which the filters are sampled from a pre-specified distribution, tend to work very well for certain tasks \cite{coates2012learning, morrow2017convolutional}.
\end{enumerate}

Though neural networks are computationally intensive to train, computing the output of pre-trained or random neural networks is relatively efficient.  Thus, restricting our attention to pre-trained and random networks allows us to easily process millions of images in the cloud.
Using PyWren \cite{jonas2017occupy}, a serverless framework for python, we can easily scale up our embarrassingly parallel featurization by using AWS Lambda and AWS EC2. With this framework we can process 1 million images in less than 1 hour, which allows us to analyze problems at large spatial extents and fine resolution.


After featurizing, we solve an L2-regularized linear regression problem to obtain model predictions. Performance is measured with respect to a held out data set sampled UAR from the state of Arizona. We report performance in terms of mean average precision (MAP), a metric that accommodates for both type 1 and type 2 error in predictions. This metric also corresponds to the area under the precision-recall curve, and is commonly used in machine learning literature.

\section{Initial results}
\begin{figure*}[!h]
\includegraphics[width=\textwidth]{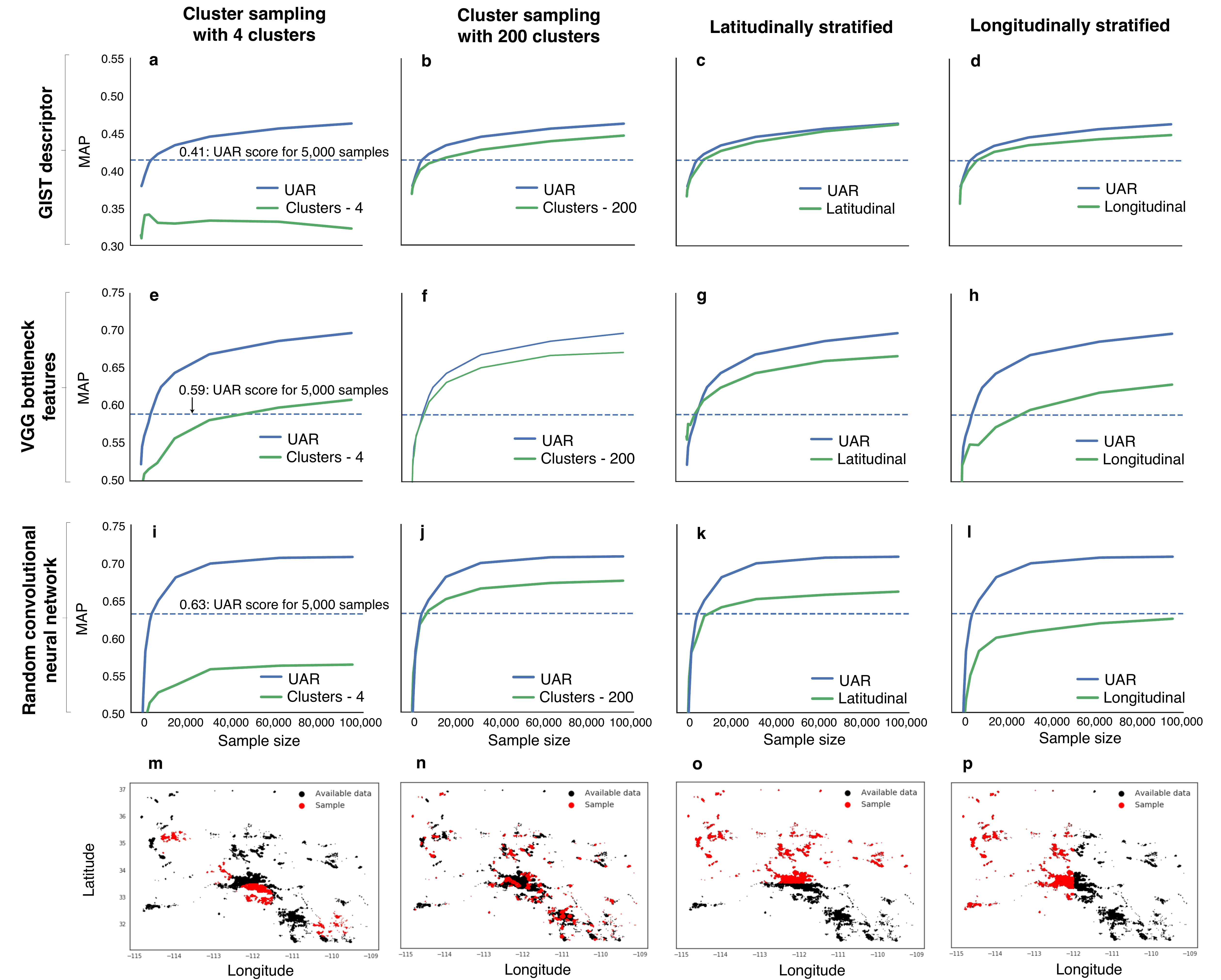}
\caption{The effects of sampling strategy, featurization technique, and sample size on the predictive power of satellite imagery. Panels \textbf{(a)} through \textbf{(d)} use a GIST descriptor; panels \textbf{(e)} through \textbf{(h)} use VGG bottleneck features; panels \textbf{(i)} through \textbf{(l)} use a random convolutional neural network. Columns show different sampling strategies. Panels \textbf{(m)} through \textbf{(p)} show the houses that were sampled (red) and all non-sampled houses in the corpus of AZ house prices (black).}
\label{fig:uar_matrix}
\end{figure*}

The results of using the sampling schemes and classification techniques defined above are shown in Fig. \ref{fig:uar_matrix}, where each sampling scheme is compared with UAR sampling. The curves show mean average precision as a function of sample size.  \footnote{Between the submission and the final version of this paper we have changed the data to better reflect the actual task at hand.  We made sure every point in our test set is at least 100 meters from anypoint in training set. This prevents any overlap between satellite images between test and train set. This change causes an overall decrease in predictive power, but maintains all claims about sampling strategy in the submission version of paper.}

Results obtained from UAR sampling (shown in blue in Figs. \ref{fig:uar_matrix}(a:l)) are consistently better than those obtained by the other sampling strategies. The dashed blue lines represent the performance of the model trained from 5,000 UAR-sampled training points. In many comparisons, other sampling strategies require many times more samples to reach the 5000-point performance of UAR. Some strategies (i.e. latitudinal) perform close to as well as UAR at low sample number, yet additional data provides a smaller marginal benefit, resulting in larger performance differences at larger sample size. In some cases, e.g. cluster sampling with 200 clusters using VGG features, we require more than 40,000 examples (more than 8x UAR examples) to reach the performance of using 5,000 UAR sampled data points. In fact, the leveling off of the curves (exemplified by the cluster sampling data with 4 clusters) indicates that there may be fundamental limitations to the prediction power we can achieve with certain strategies, and furthermore, that these asymptotes are significantly lower than the prediction power obtainable from UAR sampled data.

\begin{figure*}[!h]
\centering
\includegraphics[width=.7\textwidth]{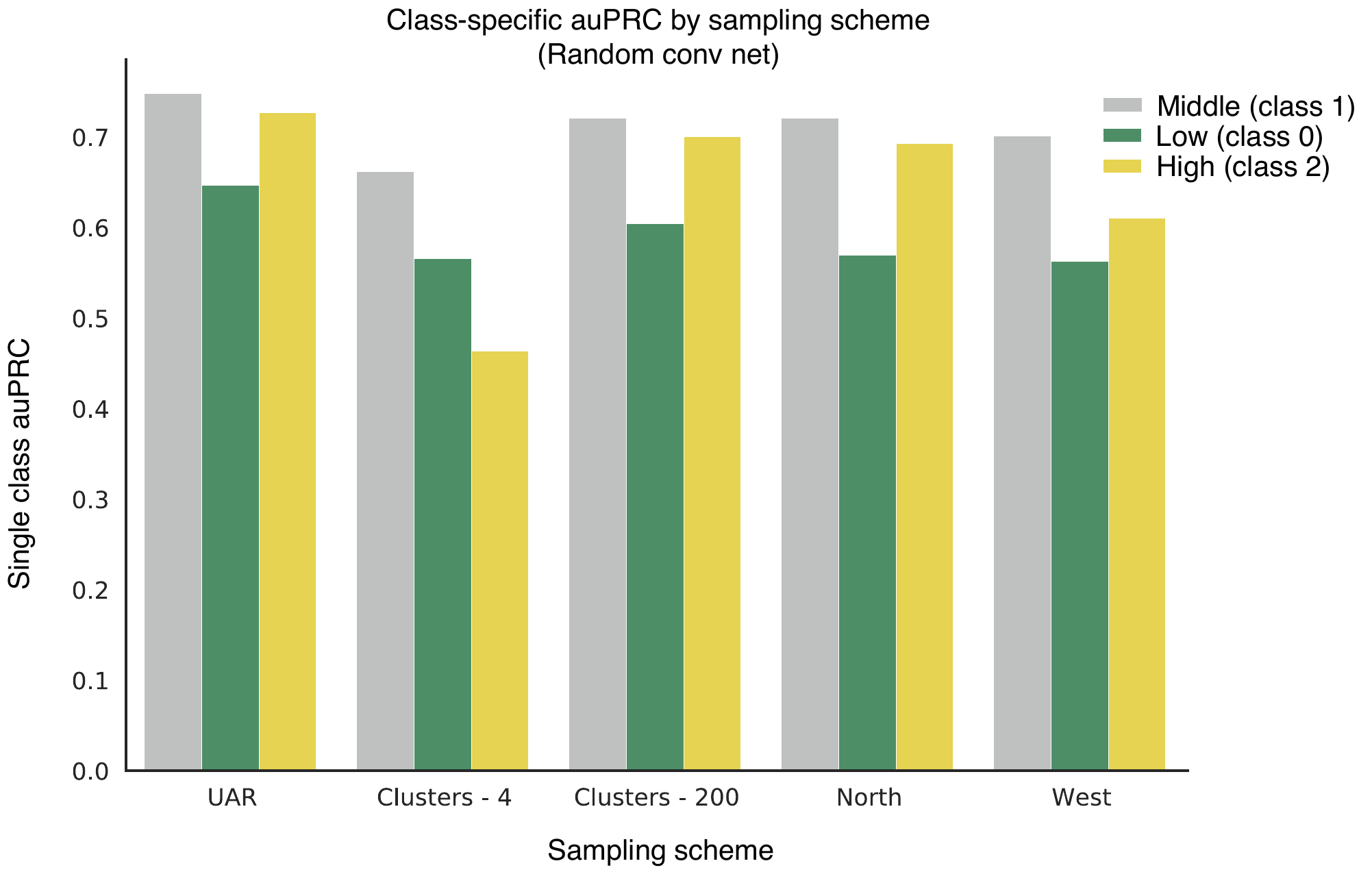}
\caption{Performance of satellite imagery classification task using a random convolutional neural network across different sampling schemes. One vs. all binary prediction performance is compared for houses in each of the three label classes: the lowest $\sigma$ of the sale price distribution (green), the highest $\sigma$ (yellow), and the remaining center of the distribution (grey).}
\label{fig:barchart}
\end{figure*}

Fig. \ref{fig:barchart} displays prediction performance broken down by class label, where 100,000 data points are sampled according to each sampling scheme.  As expected, UAR outperforms all other sampling schemes, with cluster sampling with 4 centers performing worst.  Within sampling schemes, the consistent discrepancy in performance for the middle class as opposed to the low and high classes can be explained as an effect of class imbalance in the linear solve. 

One striking feature of this breakdown of performance is that most of the variation between sampling schemes occurs within the tails of the distribution (the `Low' and `High' classes).  One compelling explanation for this trend is that sampling schemes which expose more variety of data result in higher predictive power in these smaller label classes. This would explain why UAR performs best, followed by schemes that sample over a wide spatial range (clusters with 200 centers, longitudinal, and latitudinal), and lastly followed by the scheme which has the most limited range of spatially distributed examples (clusters with 5 centers). 

In sum, these results imply that sampling scheme is an important factor to consider in imagery prediction tasks, especially when making predictions over subpopulations.

\section{Discussion}

In most machine learning problems, including image classification tasks, data are treated as if generated from an independent and identically distributed process. However, in reality, most remote sensing tasks use observations that are generated from non-random, non-uniform processes, such as cluster sampling or other forms of geographic stratification. In locations where remote sensing is uniquely valuable, cluster sampling is particularly common. For example, the DHS serves as the predominant source of health data covering developing countries, and conducts surveys using censuses of village clusters, as do the World Bank’s LSMS surveys. Non-uniform sampling is generally unavoidable in remote sensing applications -- researchers can only train models using limited available ground observations.

By sub-sampling a dataset of the all home sale prices in Arizona, we quantify the loss in predictive power suffered under non-uniform sampling schemes. Our results highlight the substantial gains from uniform random sampling, regardless of the featurization applied to the raw satellite imagery. Moreover, we demonstrate that gains resulting from increased data collection, while large, are often dwarfed by gains realizable from modifying sampling design. These preliminary findings suggest that the pragmatic benefits of cluster sampling or other geographic stratification methods (such as minimizing transportation costs) should be carefully weighed against the substantial performance costs of foregoing uniform sampling. With further testing of this model both within and outside the United States, our approach has the potential to directly inform the development of adaptive sampling strategies that can effectively combine small amounts of ground-based observation with spatially comprehensive, high resolution satellite data.

These initial findings represent a single application of a broader, scalable system for collection, storage, and aggregation of a large number of high-resolution satellite images. We plan to apply this scalable system to study the generalizability of our initial findings to other regions of the United States, and eventually to the international contexts where traditional remote sensing applications to economic development questions proliferate. In future work, we seek to create policy-relevant output that guides the development of optimal adaptive sampling strategies, where the location-specific, on-the-ground costs of implementing a given strategy are weighed against the benefits of improving predictive power. 

\section*{Acknowledgements}
This material is based upon work supported by the National Science Foundation Graduate Research Fellowship under Grant No. DGE 1106400.
\bibliographystyle{alpha}
\bibliography{prox.bib,abstract_bib_IB.bib}

\newcommand{\etalchar}[1]{$^{#1}$}
\begin{thebibliography}{TATCZ11}

\bibitem[CN12]{coates2012learning}
Adam Coates and Andrew~Y Ng.
\newblock Learning feature representations with k-means.
\newblock In {\em Neural networks: Tricks of the trade}, pages 561--580.
  Springer, 2012.

\bibitem[HAE16]{huh2016makes}
Minyoung Huh, Pulkit Agrawal, and Alexei~A Efros.
\newblock What makes imagenet good for transfer learning?
\newblock {\em arXiv preprint arXiv:1608.08614}, 2016.

\bibitem[HSW12]{henderson2012measuring}
J~Vernon Henderson, Adam Storeygard, and David~N Weil.
\newblock Measuring economic growth from outer space.
\newblock {\em American Economic Review}, 102(2):994--1028, 2012.

\bibitem[JBX{\etalchar{+}}16]{jean_combining_2016}
Neal Jean, Marshall Burke, Michael Xie, W.~Matthew Davis, David~B. Lobell, and
  Stefano Ermon.
\newblock Combining satellite imagery and machine learning to predict poverty.
\newblock {\em Science}, 353(6301):790--794, August 2016.

\bibitem[JGBH05]{jensen_using_2005}
Ryan Jensen, Jay Gatrell, Jim Boulton, and Bruce Harper.
\newblock Using {Remote} {Sensing} and {Geographic} {Information} {Systems} to
  {Study} {Urban} {Quality} of {Life} and {Urban} {Forest} {Amenities}.
\newblock {\em Ecology and Society}, 9(5), January 2005.

\bibitem[JVSR17]{jonas2017occupy}
Eric Jonas, Shivaram Venkataraman, Ion Stoica, and Benjamin Recht.
\newblock Occupy the cloud: Distributed computing for the 99\%.
\newblock {\em arXiv preprint arXiv:1702.04024}, 2017.

\bibitem[LF97]{lo_integration_1997}
C.~P. Lo and Benjamin~J. Faber.
\newblock Integration of landsat thematic mapper and census data for quality of
  life assessment.
\newblock {\em Remote Sensing of Environment}, 62(2):143--157, November 1997.

\bibitem[LW07]{li_measuring_2007}
G.~Li and Q.~Weng.
\newblock Measuring the quality of life in city of {Indianapolis} by
  integration of remote sensing and census data.
\newblock {\em International Journal of Remote Sensing}, 28(2):249--267,
  January 2007.

\bibitem[LZZ{\etalchar{+}}14]{li_review_2014}
Miao Li, Shuying Zang, Bing Zhang, Shanshan Li, and Changshan Wu.
\newblock A {Review} of {Remote} {Sensing} {Image} {Classification}
  {Techniques}: the {Role} of {Spatio}-contextual {Information}.
\newblock {\em European Journal of Remote Sensing}, 47(1):389--411, January
  2014.

\bibitem[MSP{\etalchar{+}}17]{morrow2017convolutional}
Alyssa Morrow, Vaishaal Shankar, Devin Petersohn, Anthony Joseph, Benjamin
  Recht, and Nir Yosef.
\newblock Convolutional kitchen sinks for transcription factor binding site
  prediction.
\newblock {\em arXiv preprint arXiv:1706.00125}, 2017.

\bibitem[OBLS14]{oquab2014learning}
Maxime Oquab, Leon Bottou, Ivan Laptev, and Josef Sivic.
\newblock Learning and transferring mid-level image representations using
  convolutional neural networks.
\newblock In {\em Proceedings of the IEEE conference on computer vision and
  pattern recognition}, pages 1717--1724, 2014.

\bibitem[OT01]{oliva2001modeling}
Aude Oliva and Antonio Torralba.
\newblock Modeling the shape of the scene: A holistic representation of the
  spatial envelope.
\newblock {\em International journal of computer vision}, 42(3):145--175, 2001.

\bibitem[SZ14]{simonyan2014very}
Karen Simonyan and Andrew Zisserman.
\newblock Very deep convolutional networks for large-scale image recognition.
\newblock {\em arXiv preprint arXiv:1409.1556}, 2014.

\bibitem[TATCZ11]{tapiador_deriving_2011}
Francisco~J. Tapiador, Silvania Avelar, Carlos Tavares-Corrêa, and Rainer Zah.
\newblock Deriving fine-scale socioeconomic information of urban areas using
  very high-resolution satellite imagery.
\newblock {\em International Journal of Remote Sensing}, 32(21):6437--6456,
  November 2011.

\end{thebibliography}

\end{document}